\documentclass[showpacs,amsmath,amsfonts,amssymb,pra,twocolumn]{revtex4}
\usepackage{dcolumn}
\usepackage{bm}
\usepackage[english]{babel}
\usepackage[dvips]{graphicx}
\begin{document}

\title{Non-destructive interferometric characterization of an optical dipole trap}
\author{Plamen G. Petrov}
\email{petrov@bgu.ac.il} \altaffiliation[Present address:
]{Department of Physics, Ben Gurion University, Be'er-Sheva 84105,
Israel}
\author{Daniel Oblak}
\affiliation{QUANTOP, Danish National Research Foundation Centre of
Quantum Optics, Niels Bohr Institute, DK-2100 Copenhagen \O,
Denmark.}
\author{Carlos L. \surname{Garrido Alzar}}
\altaffiliation[Present address: ]{Laboratoire Charles Fabry, CNRS
et Universit\'{e} Paris 11, 91403 Orsay, France}
\author{Niels Kj\ae rgaard}
\affiliation{QUANTOP, Danish National Research Foundation Centre of
Quantum Optics, Niels Bohr Institute, DK-2100 Copenhagen \O,
Denmark.}
\author{Eugene S. Polzik}
\affiliation{QUANTOP, Danish National Research Foundation Centre of
Quantum Optics, Niels Bohr Institute, DK-2100 Copenhagen \O,
Denmark.}

\date{\today}
\pacs{42.50.Lc, 42.50.Nn, 06.30.Ft, 03.65.Ta} \keywords{Quantum
non-demolition measurement, Spin squeezing, Atomic projection noise,
Atomic clocks}

\begin{abstract}
A method for non-destructive characterization of a dipole trapped
atomic sample is presented. It relies on a measurement of the
phase-shift imposed by cold atoms on an optical pulse that
propagates through a free space Mach-Zehnder interferometer. Using
this technique we are able to determine, with very good accuracy,
relevant trap parameters such as the atomic sample temperature, trap
oscillation frequencies and loss rates. Another important feature is
that our method is faster than conventional absorption or
fluorescence techniques, allowing the combination of high-dynamical
range measurements and a reduced number of spontaneous emission
events per atom.
\end{abstract}

\maketitle

%%%%%%%%%%%%%%%%%%%%%%%%%%%%%%%%%%%%%%%%%%%%%%%%%%%%%%%%%%%%%%%%%%%%%%%%%%%%%%%%%%%%%%%%%%%%%%

\section{Introduction}
The first demonstration of optical trapping \cite{Chu} of laser
cooled atomic samples has paved the way towards non-dissipative
traps, with long storage times, as well as optical lattices. This
relatively long lifetime is associated  with low photon scattering
rates in far-off resonant optical traps (FORT), and had made
possible to obtain all-optical Bose-Einstein condensates (BEC)
\cite{Barret}, \cite{Weber}. Optical lattices have recently been
used to form a storage medium for strontium atoms in an optical atom
clock \cite{Katori}, that allows for long interrogation times and
higher signal-to-noise ratios. On the other hand, considering the
field of quantum information and, in particular, the engineering of
quantum states, an optical dipole trap provides a medium with a
significant atomic density, a necessary condition for the optimal
creation of atomic spin squeezed states (SSS) \cite{QND_paper}.

In all of these experiments the characterization of trapped samples
is of interest. It has been shown \cite{lye} that, in the case of
dense atomic samples, non-invasive detection methods are more
powerful than the absorption imaging and fluorescence techniques
\cite{kuppens}. Non-destructive phase-contrast imaging has already
been demonstrated for sodium BEC in a magnetic trap \cite{Andrews}.
The technique of contrast enhancement by implementation of a $\pi/2$
phase-shift between scattered and imaging light is an improved
non-destructive dark-ground imaging technique \cite{Andrews1}, where
the off-resonant light is used to image the BEC cloud on a CCD. The
phase-contrast method has also been applied in a Rabi oscillation
experiment of a rubidium BEC where a superposition between two
internal states is created \cite{cornell}. The spatial heterodyne
imaging technique has been used to  record the atomic density
distribution of a dark-spot MOT in the light interference pattern
\cite{Kadlecek}. A recent work introduces the diffraction-contrast
imaging as a non-destructive measurement technique \cite{Turner} of
cold atoms.

In this paper we present a novel method of non-destructive
characterization of optically trapped atomic samples which relies on
optical phase-shift measurements in a Mach-Zehnder white-light
interferometer, with a sensitivity limited only by the quantum noise
of the probe light. The idea behind this interferometric measurement
is presented in \cite{QND_paper} as a tool for creation of squeezed
atomic states \cite{Kitagawa} via a quantum non-demolition
measurement (QND) \cite{Kuzmich}. Because of the very high bandwidth
achievable by an interferometric measurement, a system similar to
ours is projected to be used for non-destructive real-time
stabilization of a BEC~\cite{figl}. In the presented work we
demonstrate that using relatively low power and large detuning, we
are able to determine relevant trap parameters as losses,
oscillation frequency and temperature of trapped atoms in a
measurement procedure that is faster and non-destructive, than the
conventional imaging techniques.

The paper is organized as follows. We begin with a short description
of the interaction between a polarized light field and an
unpolarized atomic sample, followed by the derivation of an equation
for the phase-shift in our current experimental scheme. Then we
proceed with the theoretical analysis of the noise sources in our
free-space white-light Mach-Zehnder interferometer. In the third
section we present our experimental apparatus and the results of our
trap characterization.

%%%%%%%%%%%%%%%%%%%%%%%%%%%%%%%%%%%%%%%%%%%%%%%%%%%%%%%%%%%%%%%%%%%%%%%%%%%%%%%%%%%%%%%%%%%%%%

\section{Theoretical description}
\subsection{Phase-shift of probe light}
As it is well known, a two level unpolarized atomic system
influences the phase of a polarized optical (probe) field near a
transition between hyperfine ground and excited states. In our case,
we consider the alkali $D$ transition $J \rightarrow J'$ between
states having total electronic angular momenta $J$ and $J'$. The
complex index of refraction $n_\Delta$ imposed on light by a sample
of cold multilevel atoms is then given by~\cite{sobelman}
\begin{eqnarray}
 \label{eq:indexref}
 n_\Delta-1 =\frac{\lambda^3}{8 \pi^2}\sum_{F,F'}N_{F}S_{JFF'J'}
 \gamma \frac{\Delta_{FF'}\!+i\gamma}{\Delta_{FF'}^2\!+\gamma^2}\ ,
\end{eqnarray}
where $F$ is the total atomic ground state angular momentum,
$S_{JFF'J'}$ is the dipole transition strength of
$|J,F\rangle\rightarrow|J',F'\rangle$ and the primed quantum numbers
refer to the excited states. We have also introduced $N_F$ for the
atomic density in the level with angular momentum $F$,
$\Delta_{FF'}= \omega-\omega_{FF'}$ for the detuning of the probe
light from the $F \rightarrow F'$ transition, the atomic HWHM
linewidth $\gamma=2\pi\times2.6$ MHz and finally the wavelength
$\lambda$, assumed to be common for all transitions making up the
considered $D$ line. The real part of the above expression is
connected with the phase-shift,
$\phi_\Delta=k_{0}l\text{Re}\{n_\Delta-1\}$, imposed on the
off-resonant probe light field with wave number $k_{0}$ by the
interaction with the atomic sample of length $l$. The phase-shift is
proportional to the populations of the atomic levels of concern
\cite{QND_paper}
\begin{eqnarray}
 \phi_\Delta\!\!&=&\!\!\sum_{F,F'}\phi_F S_{JFF'J'}\!\frac{\gamma
 \Delta_{FF'}}{\Delta_{FF'}^2\!+\gamma^2}\ ,
\label{eq:phaseshift}
\end{eqnarray}
where $\phi_F=\frac{\lambda^2l N_F}{2\pi}$ denotes the maximum
phase-shift due to the atoms on level with angular momentum $F$ if
the light is near-resonant with $F\rightarrow F'$. The equation
\ref{eq:phaseshift} describes the general case where summation
happens over all hyperfine states. In our experiment we use atomic
Cs which has two hyperfine ground states with total angular momentum
$F = 3,4$. Since our probe is blue detuned with respect to the
cycling transition $6S_{1/2}(F=4)\rightarrow 6S_{3/2}(F'=5)$ the
phase-shift due to atoms on $F=3$ is negligible and only the atomic
population of the $F=4$ ground hyperfine level contributes to the
phase-shift in Eq.(\ref{eq:phaseshift}).

%%%%%%%%%%%%%%%%%%%%%%%%%%%%%%%%%%%%%%%%%%%%%%%

\subsection{Mach-Zehnder interferometer}\label{sec:int-theory}
As mentioned in the previous section, the parameter of interest is
the optical phase of the probe light. We choose to monitor the probe
light phase in a separated arms Mach-Zehnder interferometer. One of
the arms contains the atomic sample and we refer to it as the probe
arm. The other arm is the reference arm. The output of the
interferometer is detected by a pulsed balanced homodyne detection
scheme. This provides the differential photocurrent $i_- \propto
\cos(\tilde{\phi})$ between the two output arms. The phase
difference along the probe and reference arms is given by
$\tilde{\phi}=2 \pi \big(\int_{ref} n(L) dL - \int_{probe} n(L) dL
\big)/\lambda$. The phase can shift due to changes in either the
path length or refractive index in one of the arms, or changes in
the wavelength of the probe laser. Moreover, in the probe arm the
index of refraction can change due to the presence of atoms adding a
phase $\phi_{\Delta}$, which is the desired imprint of the atomic
variable onto the light variable. Therefore, we can write the total
phase-shift as $\tilde{\phi}=\phi+\phi_{\Delta}$, where the last
term is the atomic contribution while the first term describes the
phase-shift due the remaining parameters affecting the optical
path-length.

To detect the $\phi_{\Delta}$ one must have controlled influence of
the residual phase $\phi$. Hence, it is of key importance to
identify and minimize the fluctuations of the residual phase that
would otherwise mask the information carried by $\phi_{\Delta}$. A
comprehensive analysis of the interferometer noise sources was
provided in \cite{QND_paper} and here we shall only outline a few
main results.

The phase fluctuations due to classical phase noise of the laser
vanishes when the interferometer is aligned to the white light
position~\cite{Dandridge,QND_paper}. On the hand, the classical
amplitude noise of the probe light is canceled by the balanced
detection which corresponds to positioning the interferometer at a
half fringe. The only remaining sources of noise are mechanical
vibrations that limit the sensitivity of the atomic detection at low
frequencies. Nevertheless, with these procedures applied to the
system the interferometric detection is limited by the quantum noise
of the probe light, for a wide range of experimental parameters.
This shot-noise limited behavior is documented in
Sec.~\ref{sec:interfexp}

%%%%%%%%%%%%%%%%%%%%%%%%%%%%%%%%%%%%%%%%%%%%%%%

\subsection{Atomic decoherence and probe excitations}\label{sec:probeexc}

Having addressed the sensitivity of the detection we turn to an
examination of the non-destructive character of the interaction
using a convenient parameter for the rate of excitation of atoms
caused by the probe light. As before, the fact that the probe is
detuned far from the $F=3\rightarrow F'=2,3,4$ hyperfine transitions
means that the excitations predominantly befalls atoms residing in
$F=4$. These excitations of an atom are characterized by a pulse
integrated probability $p_{e,4F'}$
\begin{equation}
 p_{e,4F'} =\frac{\sigma(\Delta_{4F'})\Phi \tau}{A}\ ,
 \label{eq:defpe}
\end{equation}
where the absorption cross section for the probe is
$\sigma(\Delta_{4F'})=(\lambda^2/3\pi)\mathcal{L}(\Delta_{4F'})$,
$F'=3,4,5$, $A$ is the cross sectional area of the probe beam,
$\tau$ is the duration of the probe pulse, and the linewidth
functions are given by:
\begin{equation}
\mathcal{L}(\Delta_4)=\sum_{F'}
S_{4F'}\frac{\gamma^2}{\Delta_{4F'}^2+ \gamma^2}.
\label{eq:abscross}
\end{equation}

For typical parameters of our experiment namely,
$\tau=2~\mu\mathrm{s}$, number of photons $n=\Phi\tau=1.3\times
10^6$, probe beam waist $w_0=20~\mu\mathrm{m}$, and a blue detuning
of $\Delta_{45'}=100~\mathrm{MHz}$ we get a value of $p_e=0.04$.

It should be noted, that with the choice of probe detuning the
excitations primarily populate the $F'=5$ excited state from which
the atoms can only decay back to the $F=4$ ground state. Hence,
while the excitation does destroy any atomic coherences there is
only little change in the population distribution among the
hyperfine levels.

\begin{figure*}[t]
\begin{center}
\includegraphics[width=0.5\textwidth]{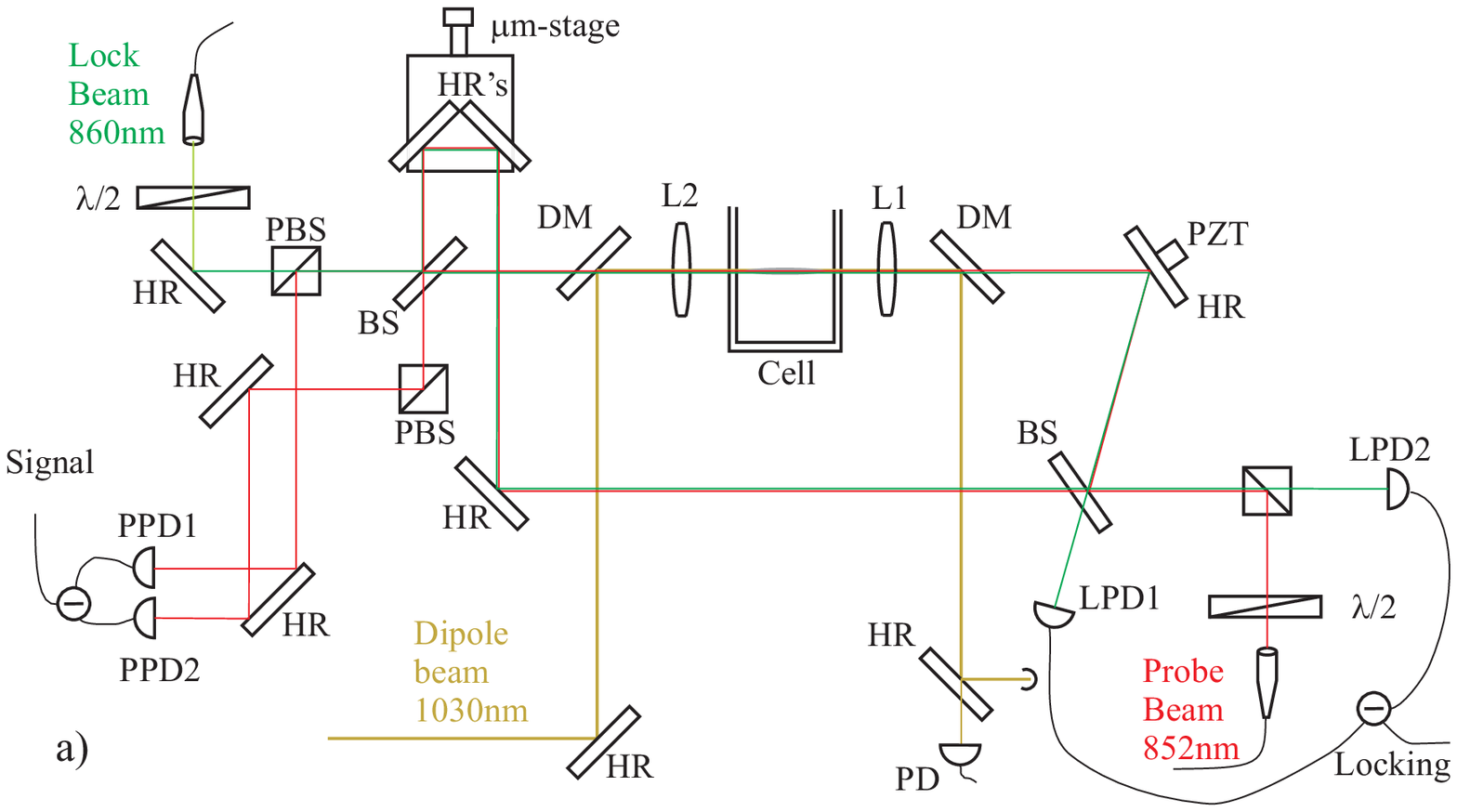}
\hspace{0.03\textwidth}%
\includegraphics[width=0.45\textwidth]{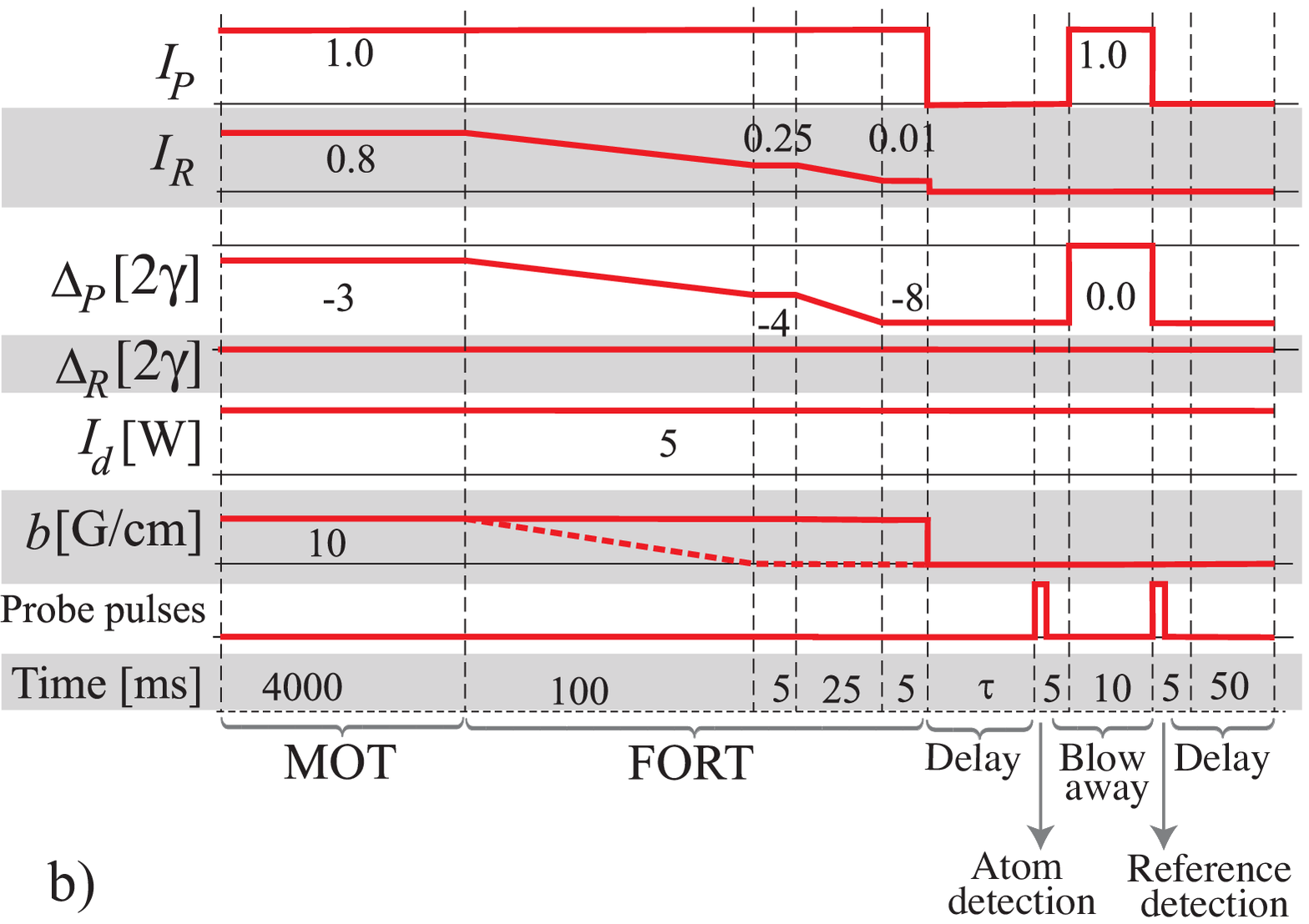}
\end{center}
\caption{(color online). (a) Sketch of the setup of the
interferometer with following elements: BS - 50/50 beam-splitter;
PBS - polarizing beam-splitter; L1 \& L2 - $f=100~\mathrm{mm}$
achromatic lenses; DM - dichroic mirrors (HT @ 852nm, HR @ 1030~nm);
PZT - Piezo electric tube; PPD1 \& -2 - Hamamatsu low noise, high
gain photodiodes for probe detection; LPD1 \& 2 - photodetectors for
locking beam detection; and half wave plates $\lambda /2$. (b)
Loading sequence of the dipole trap along with detection and
reference pulse trains. The probe pulse duration and repetition
rates are specified in the text for the different experiments}
\label{fig:interf}
\end{figure*}

%%%%%%%%%%%%%%%%%%%%%%%%%%%%%%%%%%%%%%%%%%%%%%%%%%%%%%%%%%%%%%%%%%%%%%%%%%%%%%%%%%%%%%%%%%%%%%

\section{Dipole trap characterization}

%%%%%%%%%%%%%%%%%%%%%%%%%%%%%%%%%%%%%%%%%%%%%%%

\subsection{Atomic sample preparation}\label{sec:atomic_sample}
The general loading sequence of the dipole trap is shown in
Fig.~\ref{fig:interf}(b). Initially, the atomic sample is prepared
in a standard six beam Cs magneto-optical trap (MOT). The red
detuning of the cooling laser is set to $-6 \gamma$. The number of
atoms collected in the MOT varies from $7.5\times 10^{7}$ to
$10^{8}$ depending on the partial pressure of the Cs vapor. The
typical density of the atoms in the MOT is $1.3\times 10^{9}$
atoms/cm$^{3}$.

In the next step, the atoms are further sub-doppler cooled for 150
to 200~ms. During this stage the cooling laser detuning is reduced
to $-8\gamma$ and simultaneously the intensity of the hyperfine
repump light is reduced.

Finally, the cold Cs sample is transferred into a far-off-resonant
optical dipole trap (FORT) created by a single light beam focused at
the center of the MOT. The dipole trap laser beam is generated by an
Yb:YAG laser at a wavelength of 1030~nm. The waist of the trapping
beam is estimated to be 40~$\mu$m, which along with a power of 3.5~W
amounts to a maximum optical potential depth of
$U_\mathrm{dip}/k_{B}=\mathrm{~380~\mu K}$. The typical density of
atoms in the $F=4$ state is as high as
8$\times$10$^{10}$~atoms/cm$^{3}$, which is an increase of almost 62
times compared to the MOT density. The atoms are held in the FORT
for variable times from 10 to 1000~ms before being detected with the
interferometer.

%%%%%%%%%%%%%%%%%%%%%%%%%%%%%%%%%%%%%%%%%%%%%%%

\subsection{Interferometer}\label{sec:interfexp}
The interferometer, as sketched in Fig.~\ref{fig:interf}(a), is in a
Mach-Zehnder configuration with free space propagating beams.
Compared with our earlier setup, which employed single mode fibers
\cite{QND_paper}, this interferometer has significantly lower light
losses, but is more susceptible to acoustic and vibrational
disturbances. The linearly polarized probe beam hits the
interferometer input coupler and splits 50/50 into the reference and
the probe arm. The beam in the probe arm is carefully aligned
co-linearly with the dipole trapping beam, so that the position of
the probe beam waist (20$\mu$m) overlaps with the atomic sample. The
pathlength of the reference arm can be coarsely aligned using the
mirrors mounted on a translation stage, so that the distances
covered by the probe and reference beams, when they are overlapped
on the output coupler, are roughly equal. We achieve a maximum
fringe visibility of $\mathcal{V}=98\%$, which is limited by mode
mismatching of the two interferometer arms. The pulsed probe beam is
detected with a balanced detection scheme\cite{Hansen} using the low
noise photodiodes $PPD1$ and $PPD2$. From the integral of the
differential photocurrent $i_-$ over the pulse duration, we extract
the pulse area. The mean value gives the phase-shift, and the
variance gives information about the phase fluctuations.

As in our previous work \cite{QND_paper} the Mach-Zehnder
interferometer is locked to the interference signal obtained from an
auxiliary off-resonant CW laser. The locking beam propagates through
the interferometer along the same paths as the probe beam but in the
opposite direction and with orthogonal polarization. The wavelength
and power of the locking laser are 860~nm and 1~mW, respectively. As
discussed in Sec.\ref{sec:int-theory}, to suppress the influence of
amplitude and phase noise we lock the interferometer to half fringe
at the white light position, corresponding to a nearly zero path
length difference with the help of the procedure described in
\cite{QND_paper}.

Having implemented all the measures for removing undesirable noise
contributions, the fluctuations in our interferometer signal must be
due only to the quantum noise of the probe light. As a measure of
the instability or noise of the interferometer signal we use the two
point variance $\sigma^2(\tau_0)$ defined in \cite{QND_paper}, which
describes the variation of the collected phase-shift from a sequence
of light pulses with a certain duration $\tau$ and time separation
$\tau_0$ \footnote{The time separation $\tau_{0}$ is an integer
multiple of the repetition period $T$}. In the experiment we send a
thousand probe pulses of 2$\mathrm{\mu s}$ duration with
$T=6~\mathrm{\mu s}$ repetition period. Then the integrated pulse
areas are used to compute the two point variance $\sigma^2(\tau_0)$.

\begin{figure}[t]
\begin{center}
\includegraphics[width=0.4\textwidth]{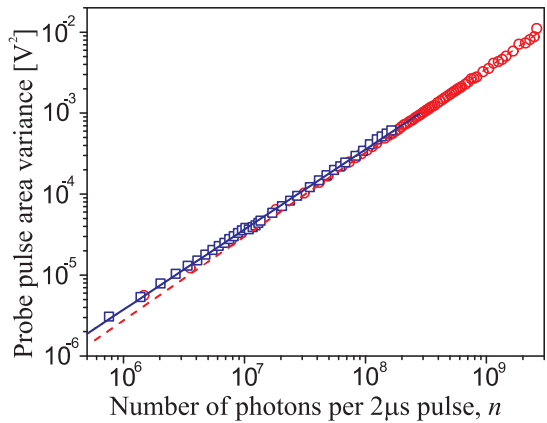}
\end{center}
\caption{(color online). Noise in the amplitude $x$ ($\circ$) and
phase $y$ ($\square$) quadratures of the probe light. Fits to the
data in log-log scale give slopes of $1.036\pm 0.008$ (red) for the
amplitude and $0.990\pm 0.005$ (blue) for the phase quadrature}
\label{fig:optnoise}
\end{figure}

The quantum shot noise of light will cause the two point variance to
grow proportionally to the probe power. On the other hand, classical
noise will result in a quadratic growth of $\sigma^2(\tau_0)$ with
the probe power. To verify the sensitivity of the balanced detection
we detect the noise in the amplitude quadrature by splitting the
light, from the reference arm only, onto the probe detectors. The
fit to the data in Fig.~\ref{fig:optnoise} shows that amplitude
noise scales linearly with the number of photons in the probe beam,
indicating that our balanced detection is shot noise limited. The
detectors remain shot noise limited on all time-scales relevant for
the measurements.

Subsequently, we unblock the probe arm, whereby the detection
becomes sensitive to phase fluctuations. The result in
Fig.~\ref{fig:optnoise} shows that the two point variance scales
linearly with the photon number. On this, we conclude that on the
6~$\mu$s time-scale the detector and interferometer sensitivities
are limited only by quantum shot noise for the probe power range
from 0.2$\mu$W - 13$\mu$W corresponding to around 2 million - 112
million photons per 2$\mu$s pulse. When the pulses have a larger
temporal separation the interferometer does not remain shot-noise
limited for high probe powers. In general the upper power at which
the interferometer is shot noise limited decreases with increasing
pulse separation. For measurement on the scale of several ms we find
that a probe power of 150 nW ensures shot-noise limited operation
and at same time fulfils the requirement that the detection be
non-destructive.

As for the general feasibility of interferometric measurements it
has been verified experimentally in a simplified setup ~\cite{figl}
that through careful design the acoustic noise can be suppressed to
below the shot noise level on minute long time-scales.

%%%%%%%%%%%%%%%%%%%%%%%%%%%%%%%%%%%%%%%%%%%%%%%

\subsection{Interferometry with dipole trapped atoms}
In this section we will present the results from the non-destructive
characterization of a dipole trapped atomic sample. Probe light is
provided by a grating stabilized laser diode and detuned by
$100\mathrm{MHz}$ from the cycling
$6S_{1/2}(F=4)\rightarrow6S_{3/2}(F'=5)$ transition in caesium. The
probe power is either 150 or 300$\mathrm{nW}$. The measurements are
done as follows: the dipole trap is loaded using the sequence
described in Sec.~\ref{sec:atomic_sample}. After a variable storage
time delay, but not shorter than 10 ms, the atoms in the dipole trap
are probed by a train of light pulses with a typical pulse duration
of 2~$\mu \mathrm{s}$ and a repetition period of 6~$\mu\mathrm{s}$
to 100~$\mu\mathrm{s}$, depending on the measurement. The number of
pulses is chosen on the purpose of the measurement, but usually
varies from 10 to 100. After probing the atoms, resonant light from
the MOT cooling beams is applied in order to remove the atoms from
the probing volume and consecutively a reference measurement is
taken using the same probe light characteristics. The first
measurement collects information about atomic sample, recorded in
the light phase, and the second is a reference with no atoms in the
probe arm that eliminates the slow drifts of the residual
interferometer phase.

In the following we present the results of this measurement method
applied to various dynamics of the FORT.

%%%%%%%%%%%%%%%%%%%%%%%%%%%%%%%%%%%%%%%%%%%%%%%

\subsubsection{Loading and loss dynamics}
Here we will apply our non-destructive measurement method to
determine the loading and loss parameters of our dipole trapped
atomic sample. A comprehensive analysis of the loading dynamics of a
FORT is done by Kuppens \textit{et al}.~\cite{kuppens}. The dynamics
of the loading process into the $F=4$ hyperfine ground state are
described by the equation:
\begin{equation}\label{eq:gen_dynamics}
\frac{dN_4}{dt}=R_0\exp\left(-\gamma_{MOT} t\right)-\Gamma_{L}
N_4-\beta_{L} N_4^2,
\end{equation}
where $\gamma_{MOT}$ is the rate at which the MOT loses atoms, $R_0$
is the loading rate of the dipole trap, $\Gamma_{L}$ is the light
assisted loss rate of the FORT, $\beta_{L}$ is the light assisted
density dependent loss rate of the FORT.

\begin{figure*}[t]
\begin{center}
\includegraphics[width=0.36\textwidth]{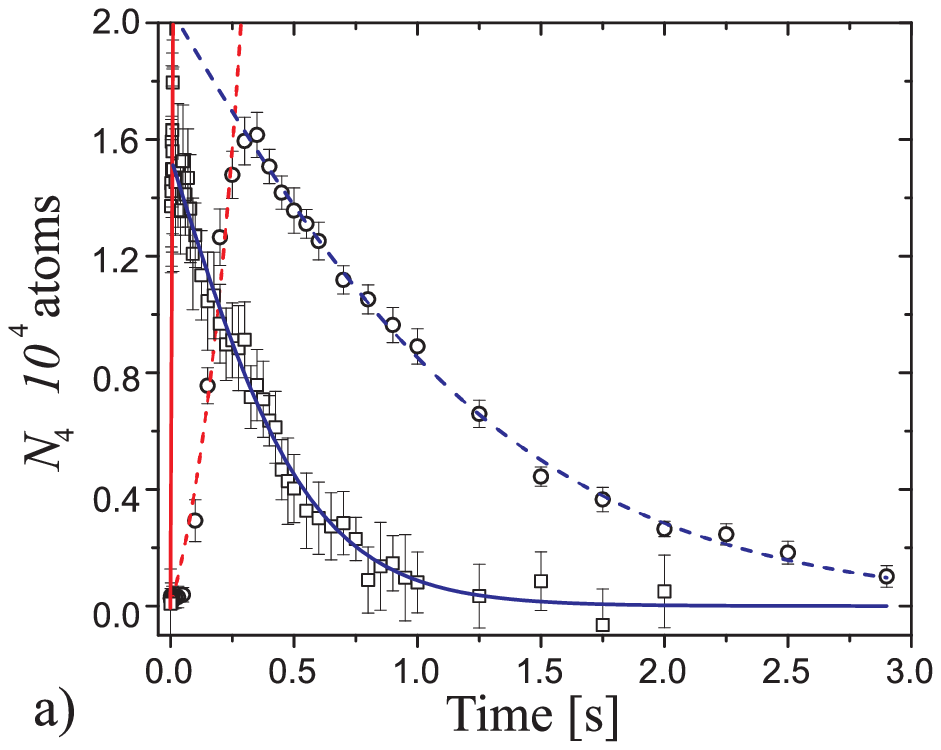}
\hspace{0.01\textwidth}%
\includegraphics[width=0.34\textwidth]{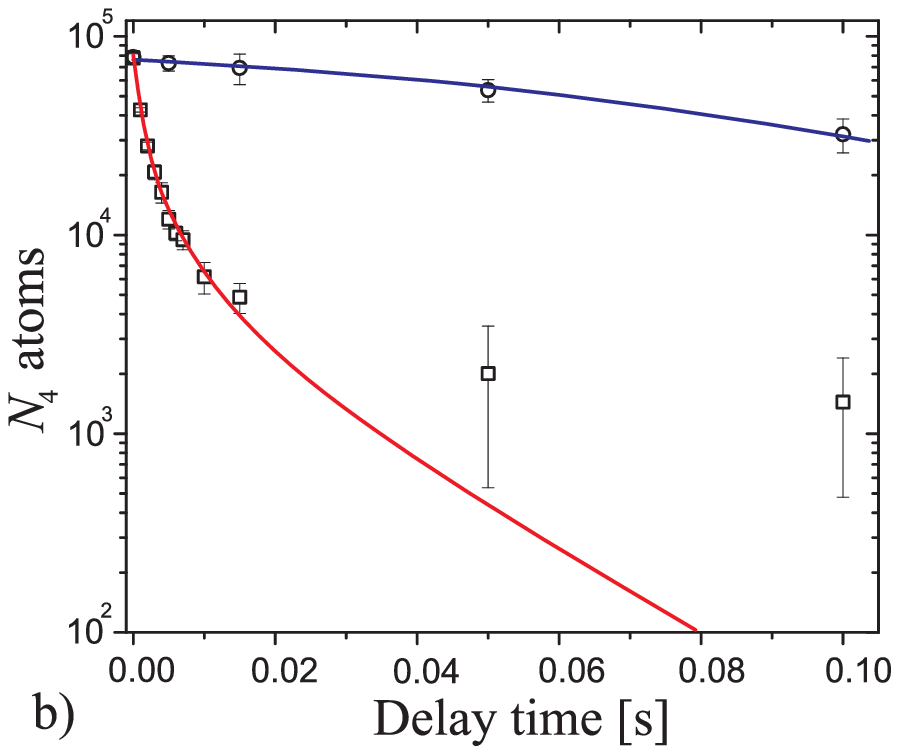}
\hspace{0.01\textwidth}%
\includegraphics[width=0.22\textwidth]{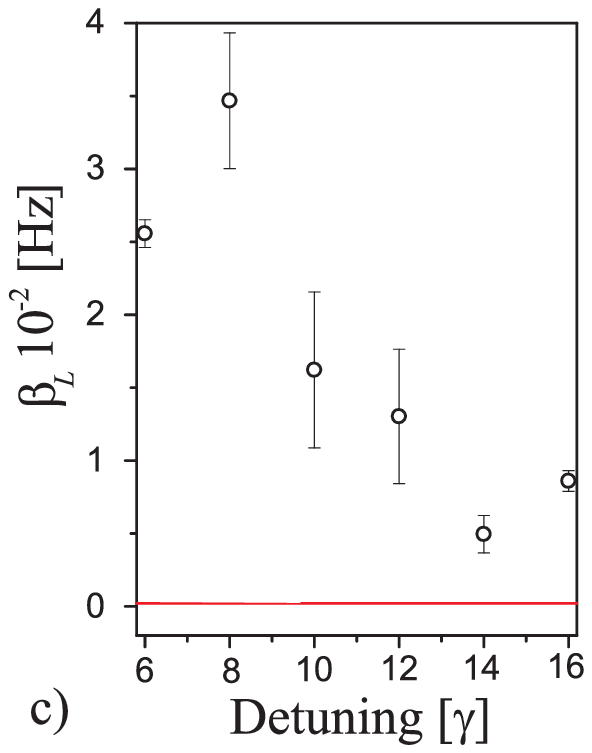}
\end{center}
\caption{(color online). Loading and loss curves of FORT: (a) Number
of atoms loaded in the FORT as a function of loading time for the
compression ($\square$) and molasses ($\circ$) regimes. The curves
are fits to different parts of Eg.(\ref{eq:gen_dynamics}). (b) Loss
curves of FORT. With MOT light on (squares, red curve). Without any
MOT light (circles, blue). (c) Light induced density dependent
losses as function of light frequency. The line shows the two body
loss level "in the dark."} \label{fig:loading}
\end{figure*}

When the loading is complete, the MOT trapping fields are switched
off and only the FORT potential remains. The evolution of the FORT
is now governed by the equation
\begin{equation}\label{eq:losses}
\frac{dN_4}{dt}=-\Gamma N_4-\beta N_4^2,
\end{equation}
which contains the light-independent loss coefficients $\Gamma$ and
$\beta$. The losses are a result of the natural decay of the dipole
trap due to collisions with the background gas particles or with
other caesium atoms, and are generally different from the
coefficients $\Gamma_{L}$ and  $\beta_{L}$, that reflect the impact
of the MOT light.

For the experimental investigation of the loading of the FORT from
the MOT we consider two regimes. In the first \textit{compression}
regime the magnetic field is gradually increased from $10
\mathrm{G/cm}$ to $12 \mathrm{G/cm}$ during the loading of the FORT,
implying that the density of the MOT increases during the loading.
At the end of the loading, the magnetic field is switched off
rapidly on a timescale of $100~\mu$s~\cite{carlos}. The detuning of
the cooling light at the beginning of loading is
$-\mathrm{6\gamma}$. In the second \textit{molasses} regime the
atoms are loaded in the FORT using optical molasses cooling. This
means that the magnetic field is switched off gradually during the
loading of the FORT (see dashed line in Fig.~\ref{fig:interf}(b)).
The detuning of MOT light is reduced to $-\mathrm{13\gamma}$.

\begin{table}
\centering
\begin{tabular}{|l|c|c|c|c|}
\hline
& \small{Comp} & \small{Mol} & \small{Light} & \small{No light}\\
\hline
$\small{R_{0}}$ & $1.34\times 10^7$ & $3.2(6)\times 10^4$ &-&-\\
$\gamma_{MOT}$ & 831 & 5 &-&-\\
$\Gamma_{L}$ & 3.5 & 1.2 & 47(20) & -\\
$\beta_{L}$ & $1.1\times10^{-4}$ & $3(1)\times 10^{-5}$ & $1.1(1)\times10^{-2}$ & -\\
$\Gamma$ & - & - & - &  21(1) \\
$\beta$ & - & - & - & $2.3(2)\times10^{-4}$\\
\hline
\end{tabular}
\caption{Loading and loss parameters of the dipole trap in
$\mathrm{Hz}$}\label{tab:FORT_load_dyn}
\end{table}

In order to determine the loading and loss rates we vary the
duration of the loading stage and measure the number of atoms that
are loaded into the FORT.  The number of atoms as a function of the
loading time for two loading schemes are presented in
Fig.~\ref{fig:loading}(a). The power of the probe beam is
$0.3~\mu\mathrm{W}$ and the detuning is set to
$\Delta_{\mathrm{45}}=\mathrm{100~MHz}$ with respect to the cycling
$F=4\rightarrow F'=5$ transition. The number of atoms is obtained
from the mean phase-shift of a train of 10 pulses with
$2~\mu\mathrm{s}$ duration and repetition period of
$40~\mu\mathrm{s}$ for the case in Fig.~\ref{fig:loading}(a)
(squares) and $10~\mu\mathrm{s}$ in the case of
Fig.~\ref{fig:loading}(a)(circles). Each point in
Fig.~\ref{fig:loading}(a) represents an average of 20 experimental
runs.

The general solution to Eq.(\ref{eq:gen_dynamics}) is expressed with
Bessel functions, which makes it involved and time consuming to fit
the experimental data on an ordinary PC. To circumvent this issue,
we have adopted a method of separating the loading process to two
parts: initial loading during which the FORT gains atoms from the
MOT, and subsequently loss of atoms caused by collisions and decay
due to cooling light. Thus, for short times we fit to a solution of
the differential equation which includes only the first term on the
right-hand-side of Eq.~\eqref{eq:gen_dynamics}, and for longer times
we fit to a solution of the Eq.~\eqref{eq:gen_dynamics} with only
the last two terms.

The values for the loss coefficients obtained from the fits in
Fig.~\ref{fig:loading}(a) are compiled in Table
\ref{tab:FORT_load_dyn}. We see that in the compression regime
loading is faster due to the magnetic field gradient, which keeps
the MOT compressed, helping atoms to be transferred to the FORT.
However, the density dependent loss-mechanism has a larger effect in
compression regime compared to the less dense molasses regime where
the $\beta_L$ coefficient is around 30 times lower. The difference
in $\Gamma_L$ between the two regimes is not significant since it
does not depend on the atomic density.

Although the above investigation provides a good characterization of
the FORT loading, the actual values for the loss coefficients
obtained are coupled to the sub-Doppler cooling used to lower the
temperature of the atoms. Specifically, the detuning of the light is
gradually increased during the loading. To study the light induced
losses independently from the loading process we use the method
suggested in \cite{kuppens}. After storing the atoms in the
dipole-trap for 10 ms we switch the MOT beams back on with total
intensity of the cooling laser of $4.6~\mathrm{mW/cm^2}$, detuning
of $-16\gamma$, and repump laser intensity of $1.1~\mathrm{mW/cm^2}$
on resonance. The light is kept on for a varying duration and when
it is switched off again the number of atoms remaining in the FORT
is measured. Each point in Fig.\ref{fig:loading}(b)(squares)
represents an average value of the number of atoms probed with 10
pulses of $\mathrm{2~\mu s}$ duration and repetition period of
$\mathrm{40~\mu s}$. The data is fitted to the solution of
Eq.(\ref{eq:losses}) [Fig.\ref{fig:loading}(b), red curve] and the
loss coefficient values obtained from the fit are listed in the
third column of Table.\ref{tab:FORT_load_dyn}. For comparison the
losses in the FORT without any MOT light are also measured and
plotted in Fig.\ref{fig:loading}(b), circles.

The data clearly shows that the FORT undergoes severe losses due to
light assisted collisions, so that the density dependent losses will
prevail over the losses due to background gas collisions. The
density independent loss rate is little affected by the presence of
the light. We perform the same measurements at a range of detunings
to determine how it affects the light assisted losses. In
Fig.~\ref{fig:loading}(c) we plot the fitted $\beta_L$ coefficients
as function of the light detuning. As expected we see that the light
assisted losses increase as the detuning is reduced.

%%%%%%%%%%%%%%%%%%%%%%%%%%%%%%%%%%%%%%%%%%%%%%%

\begin{figure}[t]
\begin{center}
\includegraphics[width=0.40\textwidth]{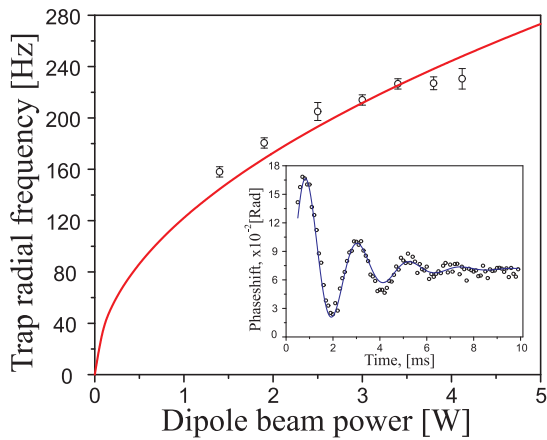}
\end{center}
\caption{(color online). Trap radial oscillation frequency as
function of dipole beam power. The estimated waist of dipole trap
beam from this fit is $\mathrm{90(1)\mu m}$. The inset: Oscillation
in the trap after its revival by switching the dipole trap laser
again after 500 $\mu s$ release. The period of the damped
oscillations corresponds to oscillation frequency of
$453(3)~\mathrm{Hz}$.} \label{fig:oscfreq_18A}
\end{figure}
\subsubsection{Oscillation frequency}

Here we adopt a different measurement technique and demonstrate that
using the interferometric measurement we can extract information
about important parameters of FORT in an measurement procedure which
is significantly less time consuming than the conventional methods.

We induce radial oscillation in the FORT by switching the Yb:YAG
trapping laser beam off for $\mathrm{500~\mu s}$ and subsequently
turning it on again. The act of re-enforcing the trap potential when
the cloud has started to expand will induce a "breathing"
oscillation of the radial size of the cloud of atoms, with a
frequency that is twice the natural radial oscillation frequency of
the trap. These oscillations are exponentially damped due to the
anharmonicity of the confining potential. We track the radial
"breathing" mode of the cloud by sending a long pulse train of 100
pulses with $\mathrm{2~\mu s}$ duration, $\mathrm{100~\mu s}$
repetition period and power of $\mathrm{150~nW}$. The result of such
a measurement is shown in the inset of Fig.~\ref{fig:oscfreq_18A}.
We emphasize that the measurement can be done in a single loading of
the FORT and that the atomic sample is not destroyed by the
measurement. All in all, the time needed for taking the data is at
maximum $\mathrm{3min}$ in the case where we average over 50 cycles.
We measured the breathing oscillations for several different values
of the dipole beam power. The result is plotted in
Fig.~\ref{fig:oscfreq_18A} and the fitted curve shows that the
oscillation frequency follows the anticipated square-root dependence
on beam power~\cite{Boiron}.

The oscillation frequency calculated from the initial estimate of
the dipole-beam radius is 2.25 times larger than the measured value.
We believe the discrepancy comes from the non-perfect spatial mode
of our trapping laser, which can be inferred from the experimentally
obtained value for the beam quality factor of $M^2=1.34$.

%%%%%%%%%%%%%%%%%%%%%%%%%%%%%%%%%%%%%%%%%%%%%%%

\begin{figure}[t]
\begin{center}
\includegraphics[width=0.4\textwidth]{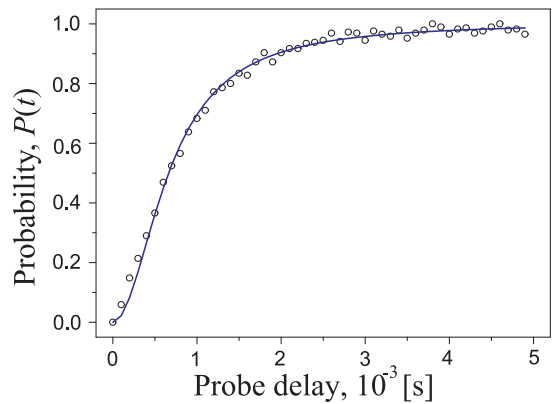}
\end{center}
\caption{(color online). Ballistic expansion of the released atomic
cloud probed in a single measurement by a train of 50 pulses,
$\mathrm{100\mu s}$ apart, with probe power of $\mathrm{150~nW}$.
The fit gives $\mathrm{T=15(2)~\mu K}$, and
$\mathrm{\nu_r=275(4)~Hz}$ respectively for the temperature and
radial oscillation frequency.} \label{fig:temperature}
\end{figure}

\subsubsection{Temperature}
Here we present our measurement of atom temperature using the
interferometer phase-shift. After storage in the dipole trap the
atoms are released by switching off the dipole laser, causing the
atomic cloud to expand ballistically in a free fall. During the free
expansion the cloud is probed by optical pulses and the measured
atomic phase-shift is used to determine the fraction of atoms that
have left the probe beam volume after a given time. In other words,
we can measure the probability $P(t)$ of an atom initially inside
the probe volume, to be outside it after a certain time. The
probability $P(t)$ is found as a convolution of the ballistically
expanding density profile and the probing Gaussian light beam. We
construct a simple model that describes the evolution of $P(t)$ as a
function of the expansion time and is similar to already known
time-of-flight techniques \cite{Weiss}. We derive an approximate
expression for the probability function which has the following
form:
\begin{eqnarray}
 \label{eq:temp_approx}
 P(t)=1-\frac{w_0^2+4\sigma_{r,0}^2}{w_0^2+4\sigma_r(t)^2}\exp\left\{-\frac{(gt^2)^2}{2[w_0^2+4\sigma_r(t)^2]}\right\},
\end{eqnarray}
where $w_0=\mathrm{20\mu m}$ is the probe beam waist,
$\sigma_r(t)=\sqrt{\sigma_{r,0}^2+\sigma_{v}^{2}t^2}$ is the
time-dependent radius of atomic sample with $g$ the earth
acceleration, $\sigma_{r,0}=\sqrt{k_BT/(M\omega_r^2)}$ and
$\sigma_{v}=\sqrt{k_BT/M}$ being the initial radial extension of the
atomic cloud and the initial atom velocity respectively. Both
$\sigma_{r,0}$ and  $\sigma_{v}$ depend on the temperature $T$,
Boltzman's constant $k_B$, the trap radial angular frequency
$\omega_r$, and $M$ the atomic mass of $^{133}\mathrm{Cs}$.

For the measurement the atoms are stored in the dipole trap for
50~ms before being released, after which probed by a pulse train
with the first pulse arriving just after the atoms are released. The
pulse train contains 50 pulses with a duration of $\mathrm{2~\mu
s}$, repetition period of $\mathrm{100~\mu s}$, and optical power of
$\mathrm{150~nW}$. The phase-shift of each pulse is normalized to
the phase-shift of the first pulse and the fraction of atoms that
have left the probing volume is calculated and plotted in
Fig.~\ref{fig:temperature}. Hence, each point on the graph in
Fig.~\ref{fig:temperature} corresponds to a single pulse and the
whole sequence of pulses belongs to a single measurement. Moreover,
a single measurement trace represents a complete ballistic expansion
of the cloud, free of shot-to-shot fluctuations in the atom number.
To correct for any possible decay due to the probe light, we take a
reference measurement with the dipole beam on and subtract it from
the ballistic data. The estimated relative decay of the detected
signal due to probe light depumping is found to be around
$\mathrm{3.5\%}$.

The results obtained for the temperature are $\mathrm{T=15(2)~\mu
K}$ and oscillation frequency $\mathrm{\nu_r=275(4)~Hz}$. The
oscillation frequency is in a reasonable agreement with the
previously obtained value of $\mathrm{244~Hz}$ (see
Fig.~\ref{fig:oscfreq_18A}) for a power of dipole laser of
approximately $\mathrm{4~W}$.

As a comparison we explore a different measurement procedure, where
the evolution of the phase-shift during the cloud expansion is
compiled from several measurements done at various points of the
expansion, similar to what would be the procedure for a destructive
measurement. With this method we obtain a temperature estimate of
$\mathrm{T=14.5(2)~\mu K}$, which agrees well with the above value.
The radial oscillation frequency estimate of
$\mathrm{\nu_r=259(4)~Hz}$ fits well with
Fig.~\ref{fig:oscfreq_18A}.

As a final verification of the results we directly measure the
expansion of the cloud radius using fluorescence imaging on a CCD
camera. By this means we get a value for the temperature of
14(2)~$\mu$K, which agrees with the interferometric measurements to
within 7~\%

\begin{figure}[t]
\begin{center}
\includegraphics[width=0.40\textwidth]{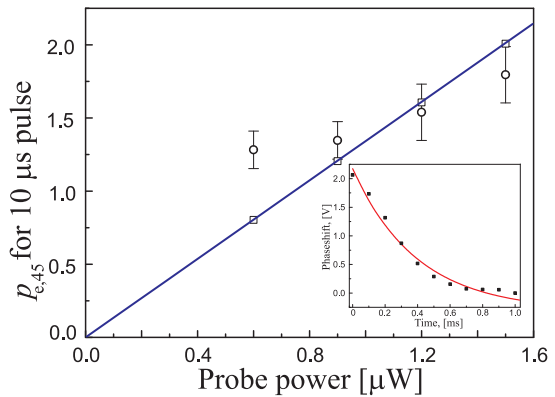}
\end{center}
\caption{(color online). Pulse integrated rate of spontaneous
emission. The experimentally obtained value for the $p_e$ parameter
as a function of the probe power for a $\mathrm{10~\mu s}$ long
probe pulses: values of $p_e$ for different power obtained from the
fit of the phase-shift decay ($\circ$), theoretically calculated
values for a probe beam waist radius of $\mathrm{21.2~\mu m}$
($\square$), and a linear fit to the data (solid line). The inset
shows the raw data for probe power of $\mathrm{1.2~\mu
W}$.}\label{fig:real_trans}
\end{figure}

%%%%%%%%%%%%%%%%%%%%%%%%%%%%%%%%%%%%%%%%%%%%%%%

\subsection{Probability of real transitions}

Here we want to verify the theoretical estimates of the
destructiveness of the measurements due to absorption of probe light
photons, which we in Sec.~\ref{sec:probeexc} chose to gauge with the
parameter $p_e$. The absorption of photons from the probe light
comes as an inevitable side-effect of the dispersive measurement.
Since the excitation is followed by spontaneous decay it will
decohere the initial atomic state and on a longer time-scale de-pump
the $F=4$ ground-level.

As the probe is blue detuned by 100~MHz from the
$\mathrm{6S_{1/2}(F=4)\rightarrow 6P_{3/2}(F'=5)}$ cycling
transition, most excitations will populate the $F'=5$ excited level
from which atoms can only decay back to the $F=4$ ground level. The
experimental method for estimating $p_e$, however, relies on the
small chance of exciting atoms to the $\mathrm{F'=4}$ excited state
whereupon the atoms may decay to either the $F=3$ or $F=4$ ground
levels. Hence, the excitation to the $F'=4$ level will cause a
reduction of population in the $F=4$ ground state that we can
observe as a reduction of the phase-shift. Since the depumping
parameter is proportional to the number of photons, we monitor the
decay of the interferometer phase-shift as a function of the probe
power. The direct experimental parameter is the characteristic time
constant of the decay that we expect to decrease linearly with
increasing probe power.

In the experiment we send a long pulse train consisting of 40 pulses
of $\mathrm{10~\mu s}$ duration and $\mathrm{100~\mu s}$ repetition
period through the atomic sample along with a reference pulse train.
Each pulse in the train causes a small amount of depumping, which
causes a smaller phase-shift of the subsequent pulse in the train.
An example of a measurement trace for a probe power of
$\mathrm{1.2~\mu W}$ is shown in the inset of
Fig.~\ref{fig:real_trans}. Altogether, the measurement is done at
four different power levels of the 100~MHz blue-detuned probe beam.

The decay is modeled by a set of rate equations that take into
account the two ground levels, the $F'=4$ and $F'=5$ excited levels
and two optical fields, namely the probe and residual hyperfine
re-pump fields. For a given probe power the value of $p_e$ is
obtained by fitting the decay of the phase-shift to the expression
obtained from the rate equations. The values of the depumping
parameter are plotted against the probe power and fitted to a
straight line as shown in Fig.~\ref{fig:real_trans}. We can
theoretically calculate $p_e$ using Eq.(\ref{eq:defpe}) and
Eq.(\ref{eq:abscross}) and if we use a probe waist radius of
$\mathrm{21.2~\mu m}$ \footnote{The value agrees with the measured
one of $\mathrm{20~\mu m}$.} the theoretical estimates coincide with
the experimental fit. We see that for very low probe powers the
pulse integrated probability of spontaneous emission per atom is
less than one. The lowest power used in our characterization
measurements is $\mathrm{150~nW}$ for a pulse duration of
$\mathrm{2~\mu s}$. This would mean that the lowest photon
scattering probability per atom is of the order of $p_e=0.038$.

%%%%%%%%%%%%%%%%%%%%%%%%%%%%%%%%%%%%%%%%%%%%%%%%%%%%%%%%%%%%%%%%%%%%%%%%%%%%%%%%%%%%%%%%%%%%%%

\section{Conclusion}
In this paper we have demonstrated a novel method for
non-destructive characterization of dipole trapped atomic sample
using a shot-noise limited free-space white-light Mach-Zehnder
interferometer.

The radial oscillation frequencies and temperature measurements were
done in a single loading run, thus enabling us for fast
characterization of atomic sample. The temperature measurement in
both single and multiple runs agrees with the one obtained by
fluorescence detection using a CCD camera. Furthermore, an
estimation of the non-destructive character of the measurement has
shown a pulse integrated probability of real transition for an atom
to be as low as 0.038.

The real time monitoring of the phase-shift in a Mach-Zehnder
interferometer allows for fast and non-destructive characterization
of atomic samples. The method is applied to a dipole trapped atomic
sample but with the same success it can be expanded to Bose-Einstein
condensates, where the optical density is much higher and the
absorption imaging does not give the necessary contrast
\cite{Andrews}. Moreover, our pulsed detection scheme allows for
microsecond timescale monitoring of processes and phenomena taking
place in an atomic cloud with a few thousand atoms, which in other
cases as absorption or fluorescence imaging are not visible due to
the limited measurement bandwidth. %requirement
%of long exposure time of the CCD cameras.

\begin{acknowledgments}

This research has been supported by the Danish National Research
Foundation and by European network CAUAC via contract
N:HPRN-CT-2000-00165.
\end{acknowledgments}

\bibliographystyle{apsrev}
\bibliography{bibfile}
\end{document}